\documentclass[letterpaper,twocolumn,10pt]{article}
\usepackage{usenix,epsfig,endnotes}
\usepackage[all=normal, title, sections, paragraphs, indent, lists, floats]{savetrees}
\usepackage{balance}			
\usepackage{booktabs}			
\usepackage{cite}			
\usepackage{courier}			
\usepackage{endnotes}
\usepackage{fancyvrb}			
\usepackage[T1]{fontenc}		
\usepackage{graphicx}			
\usepackage{listings,lstautogobble} 
\usepackage{pifont}			
\usepackage{subcaption}			
\usepackage[	usenames,
		dvipsnames,
		svgnames,
		table	]{xcolor}	
\usepackage{tikz}			
\usetikzlibrary{trees}
\usepackage{url}			
\usepackage{xspace}			
\usepackage{afterpage}
\usepackage{placeins}

\usepackage{amsthm}
\theoremstyle{definition}

\usepackage{mathtools}

\usetikzlibrary{positioning}

\graphicspath{{fig/}}
\DeclareGraphicsExtensions{.pdf}

\expandafter\def\expandafter\UrlBreaks\expandafter{\UrlBreaks%
  \do\a\do\b\do\c\do\d\do\e\do\f\do\g\do\h\do\i\do\j%
  \do\k\do\l\do\m\do\n\do\o\do\p\do\q\do\r\do\s\do\t%
  \do\u\do\v\do\w\do\x\do\y\do\z\do\A\do\B\do\C\do\D%
  \do\E\do\F\do\G\do\H\do\I\do\J\do\K\do\L\do\M\do\N%
  \do\O\do\P\do\Q\do\R\do\S\do\T\do\U\do\V\do\W\do\X%
\do\Y\do\Z\do\.\do\/\do\_\do\-\do\0\do\1\do\2\do\3\do\3\do\4\do\5\do\6\do\7\do\8\do\9}

\newif\ifcomments
\commentstrue
\ifcomments
	\newcommand{\suman}[1]{{[\textcolor{red}{#1}]}}
	\newcommand{\dimitro}[1]{{[\textcolor{green}{#1}]}}
	\newcommand{\atang}[1]{{[\textcolor{blue}{#1}]}}
	\newcommand{\theo}[1]{{[\textcolor{magenta}{#1}]}}
\else
	\newcommand{\suman}[1]{}
	\newcommand{\dimitro}[1]{}
	\newcommand{\atang}[1]{}
	\newcommand{\theo}[1]{}
\fi

\newcommand{\stexttt}[1]{{{\small{\texttt{#1}}}}}
\newcommand{\ar}[3]{} 
\include{latexdefs}

\def\eg{{e.g.,}\xspace}

\renewenvironment{thebibliography}[1]{		%
	\begin{oldthebibliography}{#1}		%
	\setlength{\parskip}{0.7ex}		%
	\setlength{\itemsep}{0.07ex}		%
	\fontsize{7.8}{8.3}
	\selectfont
}{\end{oldthebibliography}}

\lstdefinestyle{Cstyle}{
  language=C,                     
  numbersep=5pt,                  
  backgroundcolor=\color{white},  
  showspaces=false,               
  showstringspaces=false,         
  showtabs=false,                 
  tabsize=2,                      
  captionpos=b,                   
  breaklines=true,                
  breakatwhitespace=true,         
  title=\lstname,                 
  basicstyle=\scriptsize\ttfamily,
  numbers=left,                   
  xleftmargin=5.0ex,              
  keywordstyle=\color{magenta},
  commentstyle=\color[rgb]{0,0.6,0},
  stringstyle=\color{codepurple},
  numberstyle=\tiny\ttfamily\color{gray},
  rulecolor=\color{gray},
  autogobble=true
}

\date{}
\begin{document}
\title{\LARGE{Tug-of-War: Observations on Unified Content Handling}}

\author{
{\rm Theofilos Petsios, Adrian Tang, Dimitris Mitropoulos, Salvatore Stolfo, Angelos D. Keromytis, and Suman Jana} \\
    Columbia University			\\
    \{theofilos, atang, dimitro, sal, angelos, suman\}@cs.columbia.edu
}

\maketitle

\begin{abstract}
Modern applications and Operating Systems vary greatly with respect to how they
register and identify different types of content. These discrepancies lead to
exploits and inconsistencies in user experience. In this paper, we highlight
the issues arising in the modern content handling ecosystem, and examine
how the operating system can be used to achieve unified and
consistent content identification.
\end{abstract}

\section{Introduction}

Data handling lies at the very heart of modern applications. The Operating
System (OS), as well as the applications running in user space, perform complex
tasks on a multitude of file types, applying different security policies each
time.  File-associated metadata such as the Multipurpose Internet Mail
Extensions (MIMEs), the extended file attributes, or the
filename extension itself, provide \textit{indications} about the type of data
being handled. However, applications are free to perform their own processing
of files at will, without the interposition of the OS. This scheme, although
allowing for flexibility and specialized handling of app-specific content,
comes with a cost: as file parsing logic is propagated, with
variations, amongst similar applications, bugs are introduced. Such bugs have been
exploited for over a decade, resulting in the execution of active content
without the users' knowledge.

Current systems do not fully support \textit{unified} handling for different
types of content. Instead, in the majority of cases, applications are
responsible for performing their own processing or analysis on a particular
file. This may logically contradict analyses or processing performed on the
same file by the OS, or other applications. For instance, it is possible that a
given file may be treated as an image by a particular application, but as a
text file by a different application running on the same
machine~\cite{imagetragick}. More importantly, no mechanism exists between the
OS and the user space programs so that security policies or assumptions for a
particular file or file type can be updated dynamically.  For instance, despite
the fact that a PDF might be deemed malicious by a sophisticated PDF reader and
executed in a sandbox, this information is not communicated to less
sophisticated PDF readers that may be installed in the same host, and
might treat the file in a different manner.  If the file is not deleted by the
first PDF reader, it will continue to exist in the system, and the OS will be
unaware of the (malicious) characteristics of the file.
Moreover, although applications
often perform their own scanning or sniffing on a file's contents,
\textit{ignoring} the MIME or extension directives of the OS, \textit{they do not
communicate their different view of the application to the OS}.

As a result, the current content-handling ecosystem is heavily fragmented, with
different entities involved implementing different policies. In this paper, we
examine the current status of the content handling practices and how this
can be exploited by attackers or degrade user experience.  Drawing from the
weaknesses of current systems, we introduce our recommendations on how the OS can
serve as a reference oracle concerning file properties.

In particular, we propose
extending the existing extended file attributes mechanism so that,
whenever an application has a different view of a file than the OS,
this new view is registered back to the host OS and
communicated to all the other applications handling the file.
Thus, any differences in MIME type identification
amongst different applications, or between applications
and the OS, can be made known to the user.
This simple addition to file metadata does not impose backwards
compatibility limitations, but allows applications to act as cross-reference
oracles. Also, it enables the OS to detect any changes in the properties of
a file, and inform the user accordingly (\eg the user will be able to receive
a warning every time non-active is about to be executed as active or if active
content is about to be flagged as non-active).

\section{Motivating Example}
\label{sec:motivation}

Determining a file's content is not trivial, especially in cases of
\textit{polyglots}, where a file can have two different content types (\eg
a GIF that may simultaneously be a valid image and contain JavaScript).
In this case, the interpretation of the file might
solely depend on the context~\cite{mime_firefox}.
However, there are many cases in which mis-interpretation of content does not
match the user's expectations regarding the execution.

\begin{figure}[h!]
    \centering
    \includegraphics[width=0.85\columnwidth]{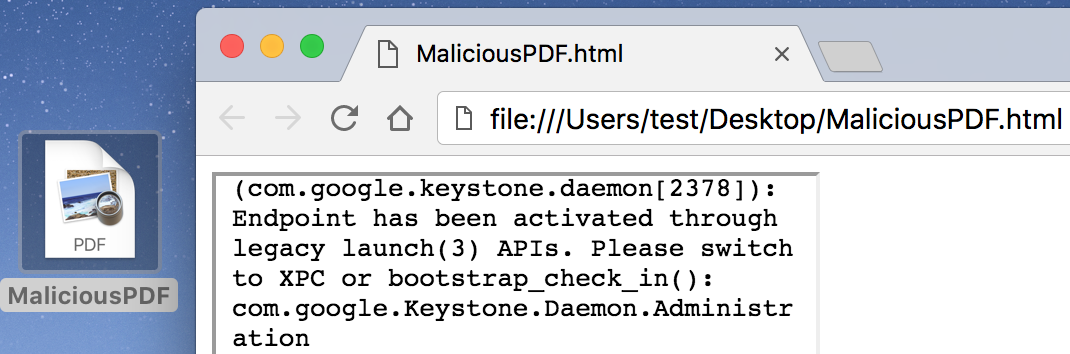}
    \caption{User's system log leaked from an HTML file executed with
    local permissions.}
    \label{fig:leak}
\end{figure}

An example of such scenario is presented in Figure~\ref{fig:leak}: Alice, a Mac
OS X user, is tricked into clicking a link that is supposed to contain a PDF
attachment. However, the website is malicious, and the attacker, instead of a
legitimate PDF, instead provides as attachment a ZIP file that contains a
Javascript executable script. The attacker has removed the extension from the
file, and has given it an icon that matches the PDF icon for Alice's OS. Once
Alice accesses the link with her Safari browser, the ZIP file is extracted into
the Downloads folder, with all the attacker's modifications on the icon and
extension preserved. Alice clicks on the PDF file, however, instead of Mac OS
X's default PDF viewer rendering the PDF, the malicious Javascript is executed
in the browser \textit{with localhost permissions}. As a result, Alice's
sensitive information such as her system logs leak to the attacker.

From the user's perspective, this is clearly unwanted behavior. This attack is
still present in today's Mac OS X El Capitan, despite the fact that similar
attacks have appeared in the past (\eg where a Javascript vulnerability allowed
loading of local files and browser-specific privileged pages into an
IFrame~\cite{CVE-2013-5598:online}). Regardless of the exploitation scenario,
Alice would perhaps have not fallen for the clickjacking attack, had the OS
enforced stricter policies with respect to maching the icon and extension of
the file with the payload to be executed, or had the OS prompt her with a
warning message.

\section{Background}
\label{sec:background}

\subsection{The Content Handling Ecosystem}
\label{subsec:echosystem}

\begin{figure}[h!]
    \centering
    \includegraphics[width=0.95\columnwidth]{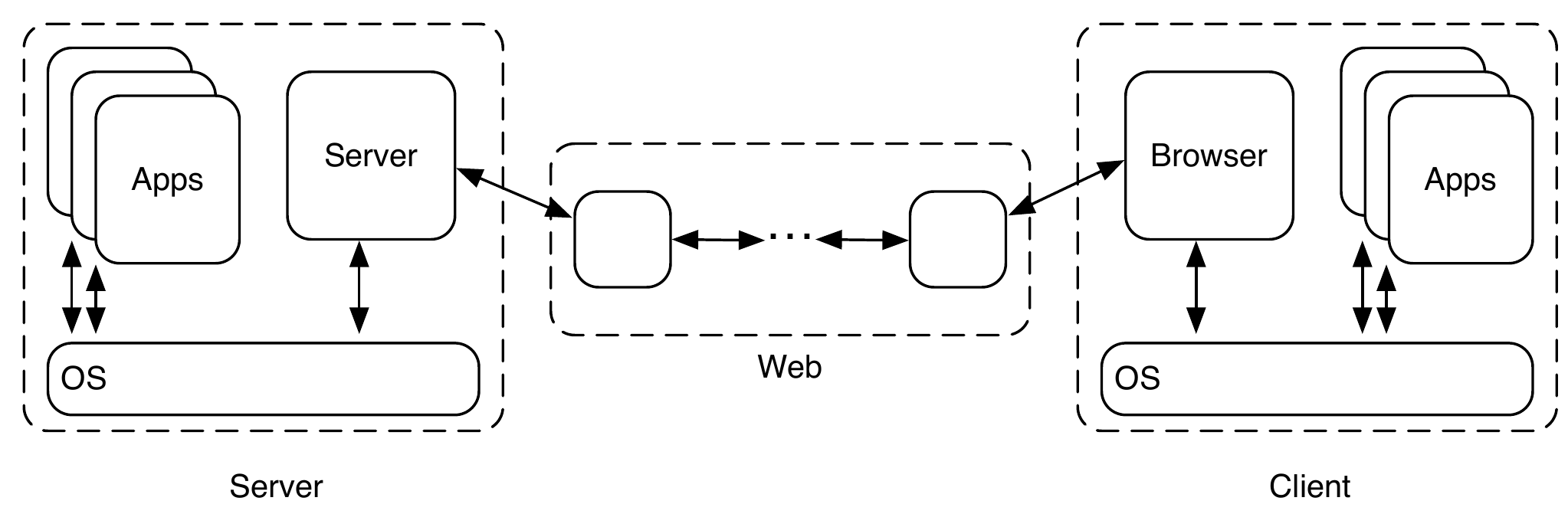}
    \caption{Content lifecycle: arrows denote content-identifying information
    exchange}
    \label{fig:lifecycle}
\end{figure}

Data are seldom processed by a single application. Instead, a file is usually
accessed by multiple programs, often running in different machines.
In a typical scenario, a payload is served from
a server and then passes through a series of
intermediate entities on the Web (\eg Ad networks, firewalls, proxies - to
end up on the client side (Fig.~\ref{fig:lifecycle}).
Once data reaches the client application, it might be stored on the
disk, passed into the OS, or it can be used as input for other applications.

Data type identification throughout this cycle is primarily achieved
through Multipurpose Internet Mail Extension (MIME) types, which characterize
the nature of a particular payload. Public MIME types should be registered with
the Internet Assigned Numbers Authority (IANA). IANA MIME types fall within nine major
categories, namely: \stexttt{application}, \stexttt{audio}, \stexttt{font},
\stexttt{image}, \stexttt{message}, \stexttt{model}, \stexttt{multipart},
\stexttt{text} and \stexttt{video}~\cite{mimetypes}. Whenever a file is
transferred from one entity to another, the MIME type of the file is usually
stored as part of the metadata of the file, sent via request headers or, left
to be determined by the target entity from scratch. At the end of this process,
a file might be written to disk to be processed in the future.  Once a file is
written on disk and it is accessed by the user, the OS is responsible to
invoke the appropriate application for each file.

\subsection{MIME handling by the OS}
\label{subsec:os}

Operating Systems associate certain file extensions with a
set of MIME types, and, likewise, applications register the MIME types they
can handle back to the host OS.
Where and how this information gets stored varies depending on the OS.~\footnote{For
instance, Ubuntu stores the mapping between file
extensions and MIME types primarily in {\stexttt{/etc/mime.types}},
Mac OS X handles MIME associations via the
{\stexttt{LaunchServices}} sub-system, whilst in Windows the respective
mappings are stored in the system Registry in
{\stexttt{HKEY\_CLASSES\_ROOT$\backslash$MIME$\backslash$Database$\backslash$Content}}.
}
If a file has a known extension the OS will open it with
the default application for the corresponding MIME type.
In the case of an extensionless file, the OS performs MIME-type sniffing
based on magic values present in the contents of the file.

\begin{figure}[h!]
    \centering
    \includegraphics[width=\columnwidth]{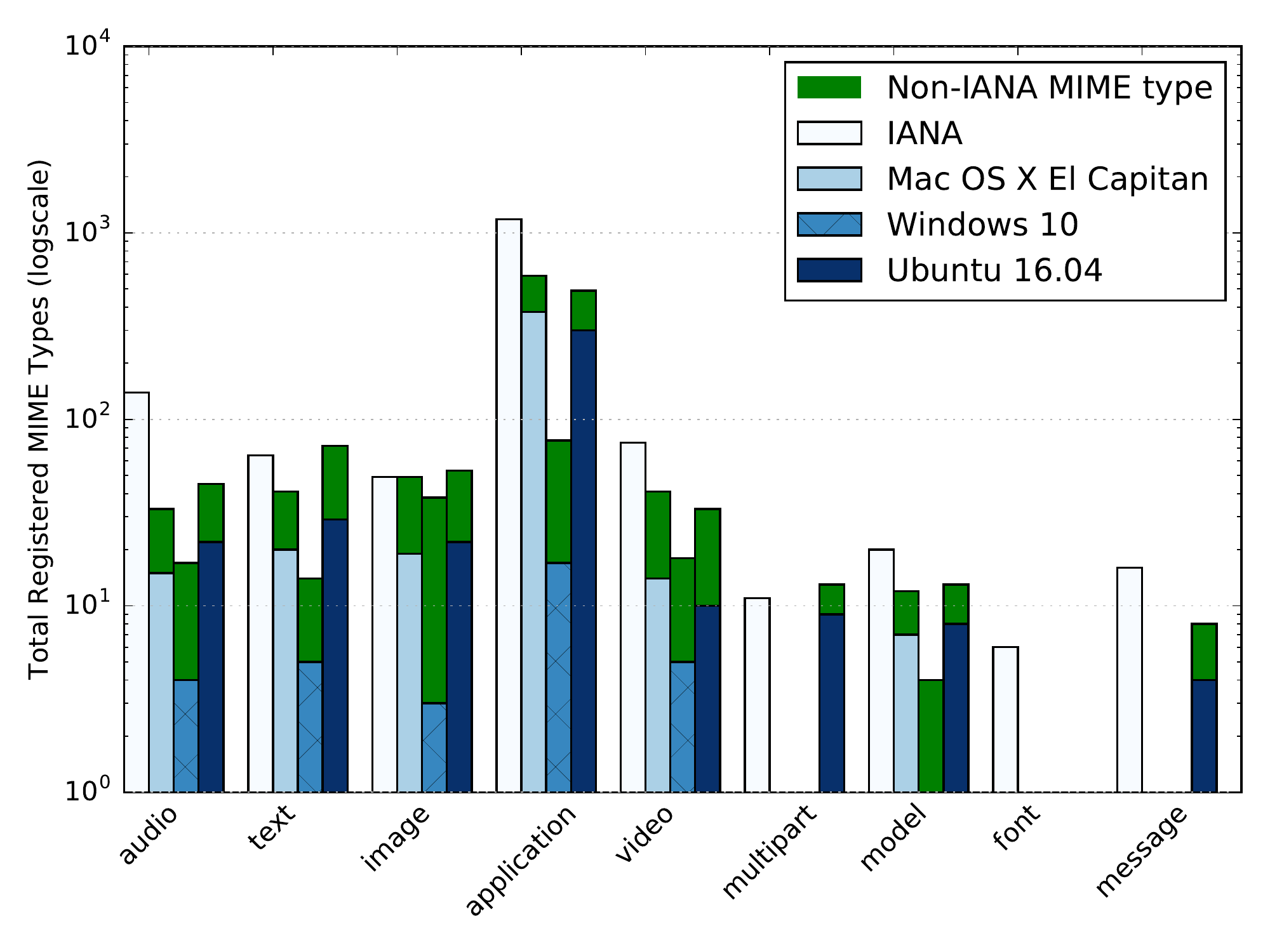}
    \caption{MIME types present in the default installation of major OSes}
    \label{fig:mime_stats}
\end{figure}

\begin{figure}[h!]
    \centering
    \includegraphics[width=0.4\columnwidth]{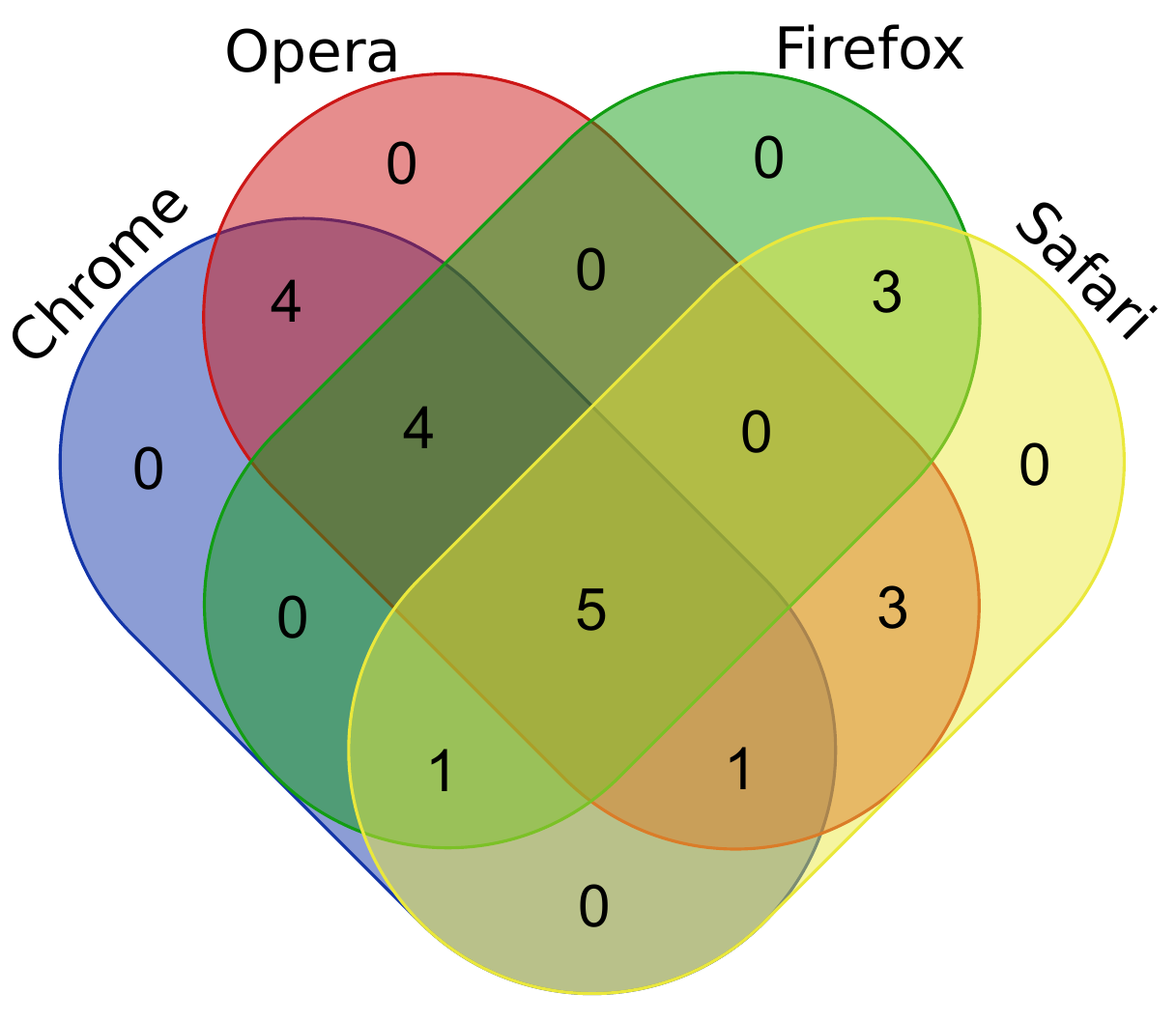}
    \caption{Venn diagram of the number of extensions handled by
    four popular browsers in Mac OS X El Capitan.}
    \label{fig:browsers}
\end{figure}

Except for the public (registered) IANA MIME types, OSes also support numerous
"unofficial" MIME types, which are used in the wild. Figure~\ref{fig:mime_stats}
shows the total MIME types register with IANA, as well as in the MIME types
present in default install
of Mac OS X El Capitan, Windows 10 and Ubuntu 16.04 distributions, per category.
We notice that, in total, the three popular OSes support an additional
Surprisingly, out of the total unique 1261 unique MIME types in the three popular
OSes, 546 MIMEs, a staggering 43.29\%, are not registered with IANA. Moreover, we
encountered 7 MIME types for El Capitan and 63 MIME types for Ubuntu 16.04 that
do not even fall within the nine major IANA categories (such instances include
MIME types that fall within the \textit{chemical, inode}, and
\textit{x-conference} families). The fact that the above numbers originate from
\textit{vanilla} systems, in which no third-party applications are installed,
is indicative of the current status in the MIME type ecosystem. Further
discrepancies are introduced if third-party applications are installed in the
system. As an example, let us consider the different extensions that are handled
by four major browsers in Mac OS X El Capitan, presented in
Fig.~\ref{fig:browsers}. We notice that despite the fact that most extensions are
handled by at least three Browsers, there are cases (\eg for filetypes such
as \stexttt{.mhtml, .webp, .mht}) for which only two of the four browsers
are registered in the OS settings.

Such instances are indicative of the underlying discrepancies between similar
applications. However, with regards to file type detection in particular,
additional complexities are introduced from the \textit{sniffing} performed
both by the OS as well as the applications, which might result in a misidentification
of a file.

\subsection{Handling Utrusted Content}
\label{subsec:ext_attr}

Although the OS stores a hardcoded MIME mapping so that different payloads can be
processed by the appropriate applications, content transferred across Operating
Systems, applications or the Web may undergo several stages of processing.
Each of these stages might modify the transferred content itself, or \textit{treat} it
as if it has a different type (\eg treat non-executable content as executable
and vice-versa). However, such assumptions or modifications are not made known
across all involved entities.  For instance, suppose that a {\small \texttt{.zip}}
file is stored on disk on server A and is being copied from server A to server
B with a MIME type of {\small \texttt{application/zip}}. If the file is
transferred from server B to an FTP server C as of type {\small
\texttt{application/x-zip}}, server C has no way of knowing that the same file
was transferred from server A to B as {\small \texttt{application/zip}}.


Thus, applications
processing untrusted files often perform their own filtering, attempting to
\textit{sniff} the type of each file they are handling. Moreover, on certain
OSes and configurations, whenever an application writes a file on disk from
an untrusted source such as the Web, it informs the OS writing metadata in the
files \textit{extended file attributes}.

Currently on Linux there are four namespaces for extended file attributes,
namely: \stexttt{user, trusted, security} and \stexttt{system}.  The
\stexttt{system} namespace
is used primarily by the kernel for access control lists (ACLs) and can only be
set by root.
Whenever a file gets downloaded from the Internet, its origin URL is stored in the
dedicated extended attribute \stexttt{user.xdg.origin.url}. If there was a
referrer URL present, the respective attribute is also set in
\stexttt{user.xdg.referrer.url}.
Windows and OS X support similar attributes. For instance, in OS X
the \stexttt{com.apple.quarantine} attribute is
set by \textit{quarantine-aware} applications to inform users about content
that originated from the Web or an untrusted source.
If an application has the \stexttt{Info.plist} key
\stexttt{LSFileQuarantineEnabled} set, all files created by that application
will be quarantined by OS X. However, although such attributes exist, not all OSes
strictly enforce properties associated with a particular file type. Thus, it
is possible for a file to have a PDF icon, without being a valid
PDF document. Such inconsistencies have been used in clickjacking
attacks~\cite{BEKBK10}, and are still present. Finally, despite the extended
file attributes being present, they are \textit{underutilized}, and are not
used to implement security policies to achieve extended access control.


\section{Observations on Current Content Handling}
\label{sec:exploits}

In this section we present some of the problems arising from the current
MIME handling ecosystem. Although MIME-related attacks are known for more
than a decade, they are still present today~\cite{imagetragick, ubuntu_crash}.


\textbf{Observation 1:}~\textit{Reliable content identification at the application
level is hard}: Relying on applications for proper content identification is not scalable,
as it is hard for all applications responsible for handling a particular file to
behave exactly the same.
To demonstrate the difficulty of
achieving a unified content policy enforcement at the application level, we
examine one of the most prevalent and well-tested categories of software: Web
browsers.
Whenever a browser attempts to identify the type of a payload, the result depends
on multiple parameters. Examples of such parameters are:
\begin{figure*}[h!]
    \centering
    \includegraphics[width=0.8\textwidth]{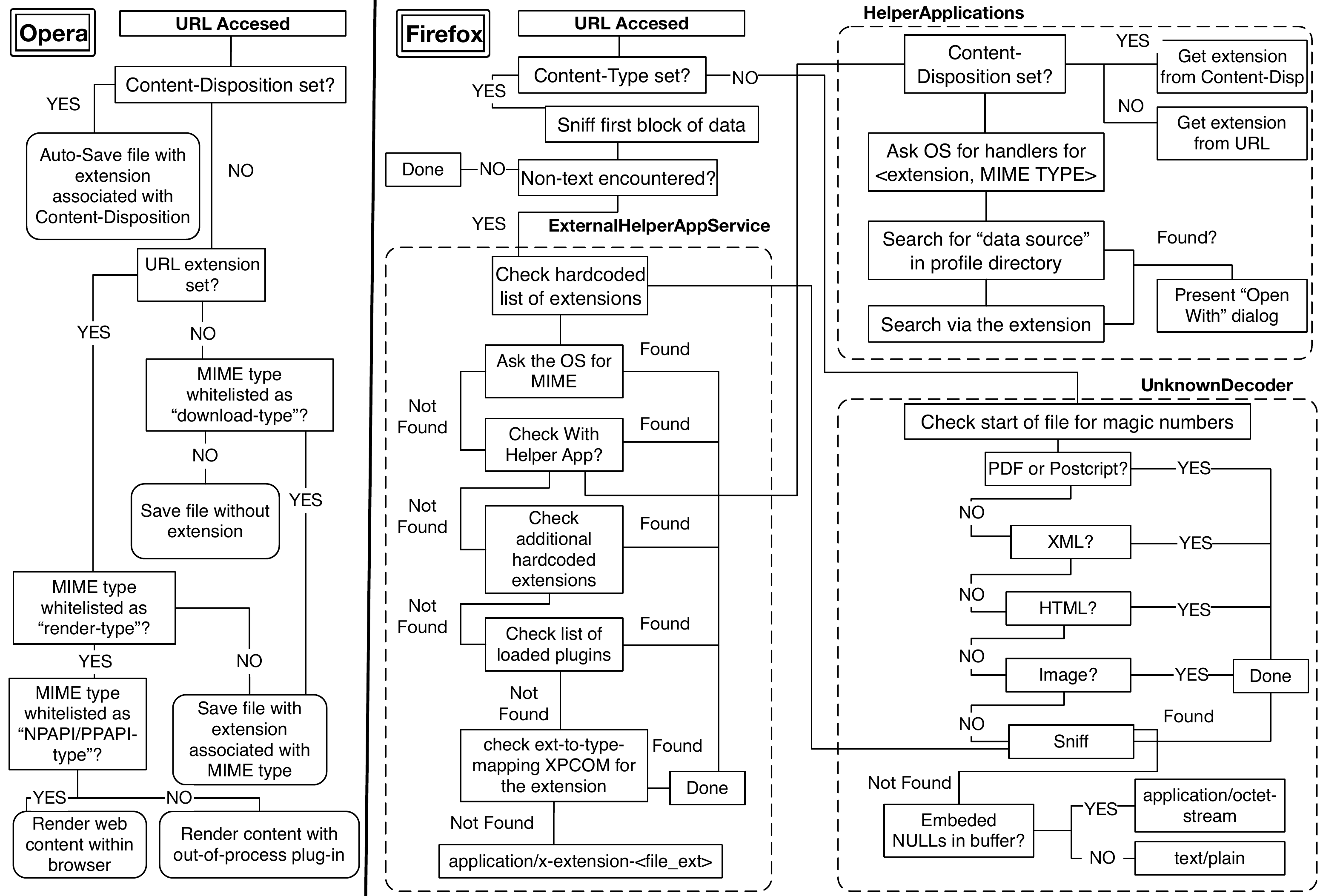}
    \caption{Discrepancies in the MIME handling of the Opera and Firefox browsers.}
    \label{fig:browser_behavior}
\end{figure*}

\begin{itemize}
    \item \textbf{Payload}: The real payload of the file being served.
    \item \textbf{Extension}: The file extension in the URL download link or
        the file extension of the payload.
    \item \textbf{Content-type}: The header sent by the server indicating
        the MIME type of the file being served. Again, this may or may not match
        the real MIME type of the file.
    \item \textbf{Content-Disposition}: The header sent by the server
        indicates whether the file should be downloaded or rendered in the
        browser.
    \item \textbf{X-Content-Type-Options}:  This header indicating whether
        the browser should skip sniffing the contents of the file.
\end{itemize}

Given the fact that there exist more than a thousand MIME
types~\cite{mimetypes}, it is almost inevitable that differences in browsers'
handling of content will continue to exist. To demonstrate the extent of the
discrepancies of modern browser, we depict in Fig.~\ref{fig:browser_behavior}
the key points in payload identification over HTTP for the Opera and Firefox
browsers. The complexity of the schemes, and the differences between only two
browsers indicate that there is high probability where a user will encounter unexpected
behavior, if the payloads are crafted properly. Web browsers are perhaps
amongst the most complex pieces of software, and browser developers often "peek"
into the implementations of other vendors to determine what functionality they
should support. Representative of the current developer workflow is the following comment,
extracted from Chromium's \stexttt{mime\_util.cc}, regarding MIME type
identification:

\textit{``We implement the same algorithm as Mozilla for mapping a file extension to a
mime type.  That is, we first check a hard-coded list (that cannot be
overridden), and then if not found there, we defer to the system registry.
Finally, we scan a secondary hard-coded list to catch types that we can
deduce but that we also want to allow the OS to override."}

Unfortunately, discrepancies between different vendors are omnipresent:
many such cases have been encountered in browser's differences in CSS and
Javascript handling. With the constant adoption of new features, such
discrepancies are prone to lead to
attacks~\cite{Keith:online,CAPECCAP82:online,crouchingtiger,facepalm}.
Not adopting a centralized scheme, not only degrades user experience with users
seeing different behavior when accessing the same resources or websites from a
different browsers, but, more importantly, leaves users vulnerable.

\textbf{Observation 2}:~\textit{Application-level content identification is risky}:
There have been multiple instances in the wild, where inactive content has been
misclassified as active and vice-versa. Such cases include input
sanitization bypasses of web-forms~\cite{CVE-2003-0130,Invision61:online,
MIMEswee24:online, CVE-2008-3823, CVE-2008-6840}, or unwanted
execution~\cite{imagetragick, ubuntu_crash}. Often exploitation is achieved
due to poor sanitization on the application side, accepting files with
fake file extensions~\cite{fake_file_ext:online}, double
extensions~\cite{CVE-2006-3102,CVE-2006-3103,CVE-2006-3104,CVE-2006-3105,
CVE-2006-2743,CVE-2006-4558} or no extensions at all. Although old, this
technique is still used today~\cite{imagetragick}.
Most of these exploitation instances however, would have been easily
detected had the OS provided a richer and reliable content identification
interface to the applications, so that the latter could delegated this task to
the Operating System.
%

Several clickjacking attacks abuse current content handling mechanisms.
For instance, Unicode character tricks have been
used so that victims think that a file has an expected, legal extension, whereas
no extension is present~\cite{packetstorm_unicode:online}. When executed, the
file will be opened according to the MIME-sniffing properties of the host OS.
Similar problems arise when users associate a particular icon with its legal
file type while the host OS allows for an arbitrary icon assignment for each
file. Even in modern OSes like the OS X, it is possible for a file to have
an icon associated with a different filetype, as we showed in our
motivating example.


\section{Recommendations for OS-based Content Policies}
\label{sec:os}

Except for instances of privilege escalation attacks within
the realm of a single application~\cite{CVE-2006-7234},
exploitation of MIME handling weaknesses often achieves
unwanted execution across applications. For instance, the \textit{auto-open}
browser feature has been exploited in the past to achieve unwanted
execution~\cite{safari_shell:online}.  Alternatively,
attackers who are aware of
the fact that browsers follow different sandboxing policies when executing local
versus remote JavaScript, might attempt to force a browser to
download a malicious payload and store it on disk, so that it executes in a
non-sandboxed environment when accessed, as was the case in our motivating example.

In the aforementioned scenarios, a discrepancy exists between either the view
of different applications on a file, or between applications and the OS.
Another common characteristic is the lack of
interfaces for communicating content-related attributes between applications
as well as between applications and the OS. This allows for a class
of attacks in which the attacker
``disguises" a payload and presents it to the appropriate application as of a
different type, so that it executes under different security policies. Since
applications often invoke other applications directly to process a particular
payload, it is necessary that they get all the possible information they could
use with regards to content identification. More importantly, \textit{any} different
view an application might have for a particular file should be communicated back
to the OS, so that the user can be warned accordingly.
In summary, it is important that the following properties are maintained:
\begin{itemize}
    \item{Each payload should be identified uniquely by all applications
        in the same OS. This uniform identification should also reflect in
        the respective properties of the payload such as the file extension
        (if any), or the icon which is associated with that payload type: All
        PDFs of a given type, should have a single icon in the system and share
        the same extension. Any discrepancies found should be reported to the user.}
    \item{Active content should be explicitly flagged based on its type, in a
        centralized manner.  Applications should let the OS know what types of
        content they consider active and file properties should be updated
        dynamically if a discrepancy is detected.}
    \item{If a user trusts a file once, this should be made known across all
        applications. Reversely, if an application detects some inconsistency
        in the type of a file that the user trusts, this should be made known
        to the user (\eg the user thinks this file is inactive but in reality
        it contains active content).}
\end{itemize}

We believe that current extended file attribute mechanism can be expanded to
provide support for the above properties. Paired with an interface for
applications to update this metadata dynamically, such a design can serve as
the basis for applying fine-grained access control at a file level. For instance,
applications may register the set of third-party executables that are
allowed to be invoked when processing a particular file.

\section{Related Work}
\label{sec:related}

Throughout Section~\ref{sec:exploits}, we refered to multiple
attacks~\cite{imagetragick, ubuntu_crash}
exploiting the current content handling ecosystem.
Except for the aforementioned examples, multiple XSS techniques have appeared
in the literature, exploiting either application flaws, discrepancies or
content mishandling~\cite{mxss_ccs13, clprotdom_sec14, barth_sp09}.
A related technique proposed by
Heiderich et al.~\cite{polyglots_ccs13}
is based on Scalable Vector Graphics (SVG).
Specifically,
the authors have illustrated that
SVG images embedded via \texttt{<img>} tag and CSS
can execute arbitrary JavaScript code.

Similar attacks utilize savvy
vectors to inject JavaScript in the web user's
browsers include Cross-channel Scripting
(XCS)~\cite{BBB09} attacks.
In an XCS attack,
the Simple Network Management Protocol (SNMP)
can be used as an attack vector.
For instance, there are several Network-Attached Storage (NAS)
devices that let web users upload files via the
Server Message Block (SMB) protocol.
An attacker could upload a file with a name that contains a
malicious script. When a benign
user connects over to the device to browse its contents,
the device will send through an HTTP response
the list of all filenames, including the malicious one.
Hence, the script that exists in this file
is going to be interpreted as legitimate by the browser.


\section{Conclusion}
\label{sec:concl}

In this work, we highlighted a number of problems arising from the currently
fragmented content handling ecosystem and possible directions that can be taken
by the OS towards resolving these issues. Content identification should not be
hard. We hope that our paper will encourage OS designers to revisit currently
deployed schemes and take steps to enable unified, cross-application content
policies.

\bibliographystyle{abbrv}
\bibliography{mimewars}

\end{document}